\documentclass{eas}
\usepackage{graphicx}
\usepackage{amsmath,amssymb}

\newcommand\vect[1]{\ensuremath{\mathbf{#1}}}

\begin{document}

\title{The solar siblings in the Gaia era} 
\author{C.A Mart\'inez-Barbosa}\address{Leiden Observatory, Leiden University, P.O.\ Box 9513 Leiden, 2300 RA, the Netherlands; \email{cmartinez@strw.leidenuniv.nl}}
\author{A.G.A. Brown}\sameaddress{1}
\author{S. Portegies Zwart}\sameaddress{1} 
\begin{abstract}
We perform realistic simulations of the Sun's birth cluster in order to predict the current distribution of solar siblings in the Galaxy.  We study the possibility of finding the solar siblings in the Gaia catalogue by using only positional and kinematic information.  We find that the number of solar siblings predicted to be observed by Gaia will be around $100$ in the most optimistic case, and that a phase space only search in the Gaia catalogue will be extremely difficult. It is therefore mandatory to combine the chemical tagging technique with phase space selection criteria in order to have any hope of finding the solar siblings.
\end{abstract}
\maketitle
\section{Introduction}
It is commonly accepted that disk stars are born in open clusters (Lada \etal\ \cite{lada}).  Particular features of the solar system suggest the Sun was born in a cluster. For instance, the radioactive elements found in the meteorite fossil record (Looney \etal\ \cite{looney}) can be explained by a supernova explosion in the vicinity of the newborn Sun. The coplanarity of the  Solar system planets  and the eccentric orbits of the Kuiper-belt objects suggest that the Sun had a close encounter with another star belonging to the same open cluster (Morbidelli \& Levinson \cite{morbidelli}). Such an encounter is expected to have occurred in the early history of the solar system (Malhotra \cite{malhotra}).    

If the Sun was born in an open cluster $4.6$~Gyr ago, the solar siblings --stars born in the same cluster- would be spread out over the Galactic disk. The identification of at least a small fraction of these stars would enable us to understand the conditions in which the solar system was formed.  

In this paper we study the evolution and disruption of the Sun's birth cluster in the Galaxy by means of state-of-the-art simulations, taking into account the effects of the spiral arms and bar of the Galaxy as well as the internal processes in the cluster. We aim to predict the current phase-space distribution of the Solar siblings   and the probability of detecting them in the large Gaia database. 

\section{Simulations set up}

The Milky Way is modelled as an analytical potential composed by an axisymmetric component together with a rotating central bar and spiral arms.  We use the potential of All\'en \& Santill\'an (\cite{allen}) to model the axisymmetric component of the Galaxy. The central bar is modelled with a Ferrers potential  (Ferrers \cite{ferrers}) and the spiral arms are modeled as perturbations of the axisymmetric potential (Antoja \etal\ \cite{antoja}).  The bar and spiral arms are non transient structures that rotate as rigid bodies with different pattern speeds. The Galactic parameters used in the simulations are taken from Mart\'inez-Barbosa \etal\ (\cite{martinezb}).

The Sun's birth cluster on the other hand is modelled as a  spherical distribution of stars that obey a Plummer potential (Plummer \cite{plummer}).  We use a Kroupa initial mass function (IMF) (Kroupa \cite{kroupa}) to model the mass distribution of the Sun's birth cluster. The minimum and maximum stellar masses are $0.08$~$M_\odot$ and $100$~$M_\odot$ respectively. Additionally, the metallicity of the Sun's birth cluster is assumed to be $Z=0.02$.  Following Portegies Zwart (\cite{simon}), we set the initial mass of the Sun's birth cluster between $500$ and $3000$~$M_\odot$. The initial radius is set between $0.5$ and $3$ parsec. 

The initial phase-space coordinates of the centre of mass of the Sun's birth cluster $(\vect{x}_\mathrm{cm}, \vect{v}_\mathrm{cm})$ are computed by integrating the orbit of the Sun backwards in time, taking into account the uncertainty in its current Galactocentric position  and velocity. By using this technique, Mart\'inez-Barbosa \etal\ (\cite{martinezb})  found the distribution of possible positions and velocities of the Sun at its birth ($P(R_\mathrm{b})$, $P(V_\mathrm{b})$). Given that for most of the Galactic parameters, ($P(R_\mathrm{b})$) is peaked around $9$~kpc, we assume initially for the Sun's birth cluster  $r_\mathrm{cm}= 9$~kpc. The Sun's birth cluster is then evolved under the influence of its self-gravity and the strong external tidal field produced by the Galaxy. We also consider stellar evolution effects. We used the \textsc{huayno} code (Pelupessy \etal\ \cite{huayno}) to compute the internal gravitational force in the Sun's birth cluster. To compute the force due to the external Galactic potential, we use a high-order Rotating \textsc{bridge} (Mart\'inez-Barbosa \etal\ \cite{martinezb}). The stellar evolution effects were  modelled with the \textsc{seba} code (Portegies Zwart \& Verbunt \cite{seba}; Toonen \etal\ \cite{toonen}).  These codes were coupled through the \textsc{amuse} framework (Portegies Zwart \etal\ \cite{amuse}). 

\section{Results}

\begin{figure}
\begin{center}
\includegraphics[width= 10 cm, height= 9cm]{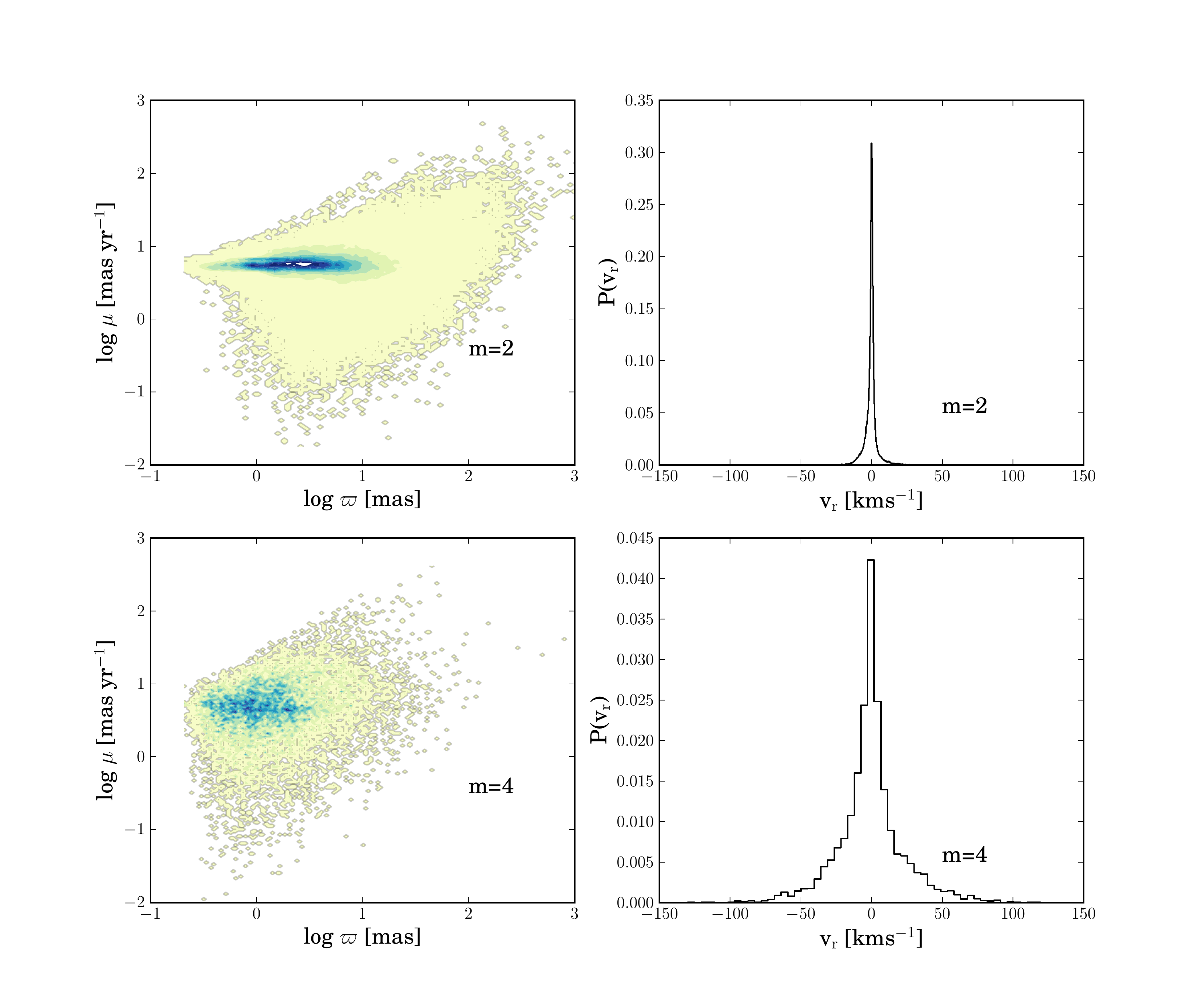} 
\caption{\textbf{Left:} Parallax Vs. proper motion of the solar siblings. \textbf{Right:} distribution of radial velocity of the solar siblings. \textbf{Top:} Astrometric properties when the Galaxy has two spiral arms. \textbf{Bottom:} Astrometric properties when the Galaxy has four spiral arms. We varied the mass and pattern speed of the bar and the amplitude and pattern speed of the spiral arms.} \label{phase}
\end{center}
\end{figure}

 Since the age of the Sun is approximately $4.6$~Gyr (Bonanno \etal\ \cite{bonano}), we evolved the Sun's birth cluster over that period of time in the Galactic model described previously. We used different initial masses and radii and different combinations of bar and spiral arms parameters. At the end of each simulation,  the Sun's birth cluster is completely disrupted, with the solar siblings located in different regions of the Galactic disk. The radial and azimuthal dispersion of the final (present day) distribution of solar siblings depends on the combination of Galactic parameters. From the galactocentric distribution of solar siblings, we compute their parallaxes, proper motions and radial velocities. An example is shown in Fig. \ref{phase}.  Here we have superposed the astrometric properties of the solar siblings for  $130$ different combinations of bar and spiral arms parameters. The initial mass and radius of the Sun's birth cluster  were fixed to $500$~$M_\odot$  and $0.5$ parsec respectively.  As can be observed, the solar siblings span a large range of parallaxes and proper motions (e.g. left panel Fig. \ref{phase}). The dark blue region corresponds to the area in phase-space where most of the solar siblings are located. The parallax of the solar siblings in this region is in the range $0.3 \lesssim \varpi \lesssim10$~mas; therefore,  most of the solar siblings are relatively far from the Sun. Note that the proper motion of the solar siblings there is around $6$~mas~yr$^{-1}$, which can be explained as stars moving along the solar circle with velocities comparable to that of the Sun (Brown \etal\ \cite{brown}).  The  distribution of radial velocities is centered around $0$~kms$^{-1}$ (e.g. right panel of Fig. \ref{phase}).  The small relative motion of the solar siblings with respect to the Sun is due to the Galactic model used in the simulations. Non-transient spiral structure does not produce significant radial migration. Transient spiral arms and  the interaction of the Sun's birth cluster with giant molecular clouds (GMCs) would produce a broader range of phase space coordinates and therefore broader radial velocity distributions. 

Even though in the simulations the Sun's birth cluster is not much affected by radial migration, we found that the expected number of solar siblings predicted to be observed by Gaia is not higher than $100$. If we assume four spiral arms in the Galactic model, this number is reduced to $50$ stars on average. Given that the Gaia catalogue will contain one billion stars, it will be extremely difficult to find  solar siblings using only phase space, even if we look for a contrast between the simulated phase-space of solar siblings and the disk stars of the Gaia data base (as done in Brown \etal\ (\cite{brown})). Therefore, in order to find possible solar siblings, it is necessary to make use of the chemical tagging technique combined with the phase space selection criteria established through the simulations.


\begin{thebibliography}{99}
\bibitem[1991]{allen} Allen, C., Santill\'an, A. 1991, Rev. Mex. Astron. Astrofis., 22, 255
\bibitem[2011]{antoja} Antoja, T., Figueras, F., Romero-G\'omez, M., Pichardo, B., Valenzuela, O.,
Moreno, E. 2011, MNRAS, 418, 1423
\bibitem[2002]{bonano} Bonanno, A., Schlattl, H., \& Patern\'o, L. 2002, A\&A, 390, 1115
\bibitem[2010]{brown} Brown, A. G. A., Portegies Zwart, S. F., Bean J. 2010, MNRAS, 407, 458
\bibitem[1877]{ferrers} Ferrers, N. M. 1877, Pure Appl. Math., 14, 1
\bibitem[1993]{lada} Lada, E. A., Strom, K. M., \& Myers, P. C. 1993,  in Protostars and Planets III,
ed. E. H. Levy \& J. I. Lunine (Tucson, AZ: Univ. Arizona Press), 245
\bibitem[2001]{kroupa} Kroupa, P. 2001, MNRAS, 322, 231
\bibitem[2008]{lindegren} Lindegren, L. \etal\  2008, in Jin W. J., Platais I., Perryman M. A. C., eds,
Proc. IAU Symp. 248, A Giant Step: From Milli- to Micro- Arcsecond Astrometry. Cambridge Univ. Press, Cambridge, p. 217
\bibitem[2006]{looney} Looney, L. W., Tobin, J. J., \& Fields, B. D. 2006, ApJ, 652, 1755
\bibitem[2008]{malhotra} Malhotra, R. 2008, in AAS/Division for Planetary Sciences Meeting Abstracts,
No. 38.01, Vol. 40
\bibitem[2015]{martinezb} Mart\'inez-Barbosa, C.A., Brown, A.G.A., Portegies Zwart, S. 2015, MNRAS, 446, 823
\bibitem[2004]{morbidelli} Morbidelli, A., \& Levison, H. F. 2004,  AJ, 128, 2564
\bibitem[2012]{huayno} Pelupessy, F. I., J\"{a}nes, J., Portegies Zwart, S. 2012, New Astronomy, 17, 711
\bibitem[1911]{plummer} Plummer, H. C. 1911, MNRAS, 71, 460
\bibitem[1996]{seba} Portegies Zwart, S. F., Verbunt, F. 1996, A\&A, 309, 179
\bibitem[2009]{simon} Portegies Zwart, S. F. 2009, ApJ, 696, L13
\bibitem[2013]{amuse} Portegies Zwart, S., McMillan, S. L. W., van Elteren, E., Pelupessy, I., de Vries, N. 2013, Computer Physics Communications, 183, 456
\bibitem[2012]{toonen} Toonen, S., Nelemans, G., Portegies Zwart, S., 2012, A\&A, 546, A70

\end{thebibliography}
\end{document}